# Solar wind reflection from the lunar surface: the view from far and near

**L. Saul, P. Wurz, A. Vorburger, and D.F. Rodríguez M.,**
Physics Institute, University of Bern, Sidlerstrasse 5, 3012 Bern, Switzerland

**S.A. Fuselier and D.J. McComas(1),**
Southwest Research Institute, San Antonio, TX 78228-0510, USA
(1) also with the University of Texas at San Antonio, San Antonio, TX 78249, USA

**E. Möbius**
Space Science Center and Department of Physics, University of New Hampshire, 39 College Road, Durham, NH 03824, USA

**S. Barabash**
The Swedish Institute of Space Physics, Box 812, SE-981 28 Kiruna, Sweden

**Herb Funsten**
Los Alamos National Laboratory, Los Alamos, New Mexico, USA.

**Paul Janzen**
University of Montana, Department of Physics & Astronomy, 32 Campus Drive, Missoula, MT 59812, USA

**Abstract:**

The Moon appears bright in the sky as a source of energetic neutral atoms (ENAs). These ENAs have recently been imaged over a broad energy range both from near the lunar surface, by India's Chandrayaan-1 mission (CH-1), and from a much more distant Earth orbit by NASA's Interstellar Boundary Explorer (IBEX) satellite. Both sets of observations have indicated that a relatively large fraction of the solar wind is reflected from the Moon as energetic neutral Hydrogen. CH-1's angular resolution over different viewing angles of the lunar surface has enabled measurement of the emission as a function of angle. IBEX in contrast views not just a swath but a whole quadrant of the Moon as effectively a single pixel, as it subtends even at the closest approach no more than a few degrees on the sky. Here we use the scattering function measured by CH-1 to model global lunar ENA emission and combine these with IBEX observations. The deduced global reflection is modestly larger (by a factor of 1.25) when the angular scattering function is included. This provides a slightly updated IBEX estimate of $A_H = 0.11 \pm 0.06$ for the global neutralized albedo, which is ~25% larger than the previous values of $0.09 \pm 0.05$, based on an assumed uniform scattering distribution.

1) **Observing Lunar Energetic Neutral Atoms**

The Interstellar Boundary Explorer (IBEX) is a NASA small explorer mission with two neutral atom cameras on a dedicated mission to explore the boundaries of heliosphere [McComas et al., 2009a]; see McComas et al. [2011] for a recent review of the early science results. The first observations of lunar ENAs [McComas et al.,

2009b] were made using the high energy sensor, IBEX-Hi sensor [Funsten et al., 2009]. Subsequently lunar ENA observations were extended to lower energies [Rodriguez et al., 2012] using data from the low energy sensor, IBEX-Lo [Fuselier et al., 2009]. The IBEX-LO sensor measures neutral atoms from 10eV to 2keV, in 8 logarithmically spaced energy bins, fully covering the range of reflected and neutralized solar wind during all but the fastest solar wind conditions.

In contrast, the Chandrayaan-1 (CH-1) mission is dedicated to lunar observation [*Goswami and Annadurai,* 2009].  IBEX and CH-1 have different types of neutral particle instrumentation, both of which enable the measurement of neutral atom fluxes.  The SARA (Sub-keV Atom Reflecting Analyzer) [Bhardwaj et al. 2005, Barabash et al. 2009] instrument on CH-1 has sensors for ions and neutrals that have multiple simultaneous separate viewing angles (pixels) while IBEX has single pixel cameras that are scanned across the sky by the spacecraft spin and motion.  Being on a lunar mission in close lunar orbit, the CH-1 instruments were designed for detailed studies of relative differences on the lunar surface, for example detecting magnetic anomaly regions and locally higher reflectivity [Wieser et al., 2009].  Images taken at different viewing angles with respect to the incoming solar wind allowed determination of the lunar scattering function.

2) **Scattering Function**

The two SARA sensors: SWIM (Solar WInd Monitor) [McCann et al. 2007] and CENA (Chandrayaan-1 Energetic Neutral Analyzer) [Kazama et al. 2007] simultaneously measure the solar wind ions, which impinge on the lunar surface, and the neutral atoms, which are backscattered or sputtered off. Both sensors include several angular sectors that are mounted with constant azimuthal angular separation. Due to this arrangement, for each measurement there is a specific pair of observation angles (azimuth and elevation) and a specific solar zenith angle.  By comparing the relative fluxes at different observation and solar zenith angles a scattering function was obtained [Schaufelberger et al., 2011].  Notation from that paper is used herein. In general, the fraction of particles leaving an observation point into some solid angle is a function of the both the position of that emission angle (theta, phi) as well as the angle with which the incoming solar wind hits the surface, the solar zenith angle (SZA).  Schaufelberger et al. [2011] found an empirical formula to describe the scattering based on three components, effectively spherical harmonic moments in the scattering angles.

3) **Viewing Geometry**

The geometry of the IBEX lunar viewing simplifies the integration of the scattering function.  The IBEX sensors are mounted perpendicular to the spacecraft spin axis, which is regularly maintained to be within ~7º sunward pointing.  This means that the Moon is always nearly a "half moon" when viewed from IBEX.  The geometry is such that for every point on the Moon illuminated by the solar wind and visible to IBEX there is only a single set of scattering parameters, theta, phi, and SZA, which determine the differential flux towards IBEX (see Figure 1).  Here theta is the polar coordinate of scattering, phi the azimuth of scattering, and SZA the solar zenith angle of scattering.

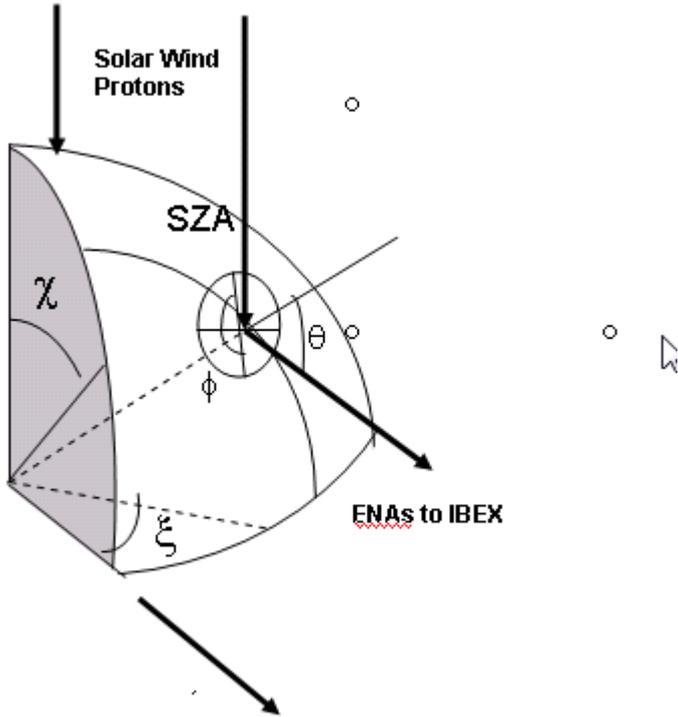

*Figure 1: The geometry of a 1/8 slice of the lunar surface is shown with the directions to IBEX and the Sun indicated. The coordinates $\chi, \xi$ describe the position on the Moon while the angles $\theta, \varphi, SZA$ describe the scattering parameters at a given position. Line normal to the surface, passing through the center of the Moon is shown, as well as an arc on the surface of constant polar scattering angle theta.*

The solar-wind illuminated portion of the Moon visible from IBEX is approximately a ¼ sphere. The scattering function itself is in general dependent on position on the Moon, due to differing properties of the materials in the regolith and also electromagnetic effects. However for IBEX observations we assume an average scattering function for the global Moon as is done in [Schaufelberger et al., 2011]. A further symmetry is immediately obvious in that this ¼ sphere is composed of two identical halves. Thus we can completely describe the scattered ENAs by considering the neutrals from a 1/8 sphere and then multiplying by a factor of 2.

Thus the differential flux towards IBEX from the 1/4 sphere will be

$$dJ = 2 \int_{\xi=0}^{\pi/2} \int_{\chi=0}^{\pi/2} f(\theta, \varphi, SZA) \sin \chi \, d\chi \, d\xi \qquad (1)$$

Where the scattering angles theta, phi, and the solar zenith angle are determined from the IBEX observation geometry to be functions of positional angles $\chi, \xi$:

$$SZA = \chi$$
$$\theta = \arcsin(\sin \chi \cos \xi) \qquad (2)$$
$$\phi = 180 - \xi$$

In making the assignments (s) we have used $r_{IBEX} \gg r_{MOON}$. For clarity a line of constant polar scattering angle $\theta$ is drawn through the scattering point in igure 1:, while the position of the scattering point on the lunar surface is $\chi, \xi$.

4) **Uniform and Cosine Scattering Cases: Normalization**

If the scattering function is uniform, then the emitted flux is equal in all directions (independent of theta and phi). For the case of cosine scattering, motivated by Lambert's law of optical scattering, there is an additional dependence on the cosine of the polar scattering angle, and we have thus defined two separate scattering functions for the uniform and cosine cases:

$$f_U(\theta, \varphi, SZA) = f_{U0} \cos(SZA)$$
$$f_C(\theta, \varphi, SZA) = f_{C0} \cos(\theta)\cos(SZA)$$
(3)

Here the dependence on the cosine of the solar zenith angle due to the lower amount of solar wind flux per unit surface area closer to the dawn or dusk line (illumination effect). Note that this cosine of solar zenith angle dependence is included in the Schaufelberger [2011] scattering function $f_S(\theta, \varphi, SZA)$.

We normalize these scattering functions such that the total number flux from a 1/4 section of the Moon is equal for each model. The total number flux can be calculated by integrating over all scattered particles for each point on the lunar section and then integrating over the lunar surface section:

$$N = 2 \int_{\xi=0}^{\pi/2} \int_{\chi=0}^{\pi/2} \int_{\varphi=0}^{2\pi} \int_{\theta=0}^{\pi/2} f(\theta, \varphi, \chi) \sin\theta \, d\theta \, d\varphi \sin\chi \, d\chi \, d\xi$$
(4)

For the case of a uniform distribution this becomes:

$$N_U = 2 \int_{\xi=0}^{\pi/2} \int_{\chi=0}^{\pi/2} \int_{\varphi=0}^{2\pi} \int_{\theta=0}^{\pi/2} f_{0U} \cos\chi \sin\theta \sin\chi \, d\theta \, d\varphi \, d\chi \, d\xi = \pi^2 f_{0U}$$
(5)

For the case of a cosine scattering function we have

$$N_C = 2 \int_{\xi=0}^{\pi/2} \int_{\chi=0}^{\pi/2} \int_{\varphi=0}^{2\pi} \int_{\theta=0}^{\pi/2} f_{0C} \cos\theta \cos\chi \sin\theta \sin\chi \, d\theta \, d\varphi \, d\chi \, d\xi = \frac{f_{0C} \pi^2}{2}$$
(6)

We perform a numeric integration of equation (4) with constants as defined in [Schaufelberger et al., 2011], which normalize to a 15° section of CH-1 viewing. For the full scattering function Eq. (4) gives $N_S = 0.0816$. We normalize our other scattering functions to this result, setting $f_{0U} = 0.0166$ and $f_{0C} = 0.0332$.

5) **Differential flux at IBEX**

We can now apply equation 1 to determine for each of our models the differential flux at IBEX and thus determine how the scattering function affects the global lunar ENA observations of IBEX. The procedure is to sum over every point on the lunar surface, the differential flux at that point that emerges towards IBEX.

For the uniform case, the total differential flux towards IBEX will be (Eq. 1)

$$dJ_U = 2 \int_{\xi=0}^{\pi/2} \int_{\chi=0}^{\pi/2} f_0 \cos \chi \sin \chi \, d\chi \, d\xi = f_{0U} \frac{\pi}{2} = 0.0130 \tag{7}$$

For the cosine case:

$$dJ_C = 2 \int_{\xi=0}^{\pi/2} \int_{\chi=0}^{\pi/2} f_{0C} \cos(\arcsin(\sin \chi \cos \xi)) \cos \chi \sin \chi \, d\chi \, d\xi = 4/3 * f_{0C} = 0.0222 \tag{8}$$

And for the full scattering function:

$$dJ_S = 2 \int_{\xi=0}^{\pi/2} \int_{\chi=0}^{\pi/2} f_S(\arcsin(\sin \chi \cos \xi), 180 - \xi, \chi) \sin \chi \, d\chi \, d\xi = 0.0104 \tag{9}$$

To compare these numbers bear in mind that the models are normalized so that they all predict the same global ENA emission, and thus global solar wind albedo. Therefore, we can see that the IBEX determination of the global neutral albedo is dependent on the scattering model.

|  | Normalization Constant | Diff. Flux at IBEX |
|---|---|---|
| Uniform Distribution | 0.0332 | 0.0130 |
| Cosine Distribution | 0.0166 | 0.0222 |
| CH-1 Determined Distribution | 1 | 0.0104 |

*Table 1: The results of the integrations from the text are shown. The normalization equates the total ENA output of the three models, thus the difference between the models in IBEX Diff. Flux (column 3) shows the dependence on scattering model.*

The uniform distribution was assumed in McComas et al. [2009] as well as Rodriguez et al. [2012]. We can see that the new scattering function thus increases the IBEX measurement of global albedo by a factor of 1.25. If the lunar emission obeyed a cosine distribution, the interpretation of a distant measurement from IBEX would be that the global lunar emission was nearly a factor of two larger.

**Implications**

Rodriguez et al. [2012] analyzed in detail six orbits when the observation of the lunar ENAs was favorable, and the observational data are given in able 2:. For four of the six orbits the correction factor is 1.25 because the observation geometry was nominal, as discussed above, during the registration of the lunar ENA signal. However, for two observations the Moon was in the magnetosheath of the Earth and thus the "illumination" by protons was somewhat different, which influences the correction factor. The difference results from two effects. The first is due to the differing portion of surface of the Moon illuminated (the Moon is no longer exactly a "half Moon" as seen from IBEX). The second is that the solar zenith angle at a given point of the surface is no longer equal to the coordinate $\chi$ ($\chi$, $\xi$ are defined with respect to the spacecraft) but rather is a larger value given by

$$SZA = \chi + \alpha \tag{10}$$

where $\alpha$ is the angular offset from the nominal viewing geometry. This second correction must then be used when doing the calculation (9), and increases the expected flux at IBEX for a given reflectivity. We find that in this geometry the determined albedo using the CH-1 scattering function appears closer to that of a uniform distribution. For a uniform scattering model there is only the illumination effect to consider, which was already taken into account by [Rodriguez et al., 2012]. The resulting correction factors per orbit are given in able 2:. Corrections when the spacecraft is inside the magnetosphere are provided by the plasma velocity model of BATS-R-US, see e.g. [Toth et al., 2011].

| IBEX Orbit Number | SOHO | | | BATS-R-US | | | IBEX $\phi_{sw}$ [°] | CH-1 |
|---|---|---|---|---|---|---|---|---|
| | Energy [eV] | Energy Standard deviation | $V_{sw}$ [km/s] | Energy [eV] | Energy Standard deviation | V [km/s] | | Scattering function correction factor |
| 29 | 783.7 | 66.52 | 387.31 | - | - | - | ~90° | 1.25 |
| 43 | 836.3 | 78.57 | 400.35 | 790.61 | 3.92 | 389 | ~75.4° | 1.06 |
| 44 | 489.6 | 37.11 | 306.32 | 659.69 | 4.03 | 356 | ~78.0° | 1.01 |
| 47 | 527.1 | 32.18 | 317.86 | - | - | - | ~90° | 1.25 |
| 58 | 649.0 | 89.37 | 352.69 | - | - | - | ~90° | 1.25 |
| 72 | 1634.8 | 35.76 | 559.74 | - | - | - | ~90° | 1.25 |

*Table 2: Shows the average values of some plasma parameters used in the simulation with BATS-R-US for the orbits of IBEX in comparison to data collected with SOHO, and shows the angle ϕ$_{sw}$ of incidence (data from Rodriguez et al., 2012).*

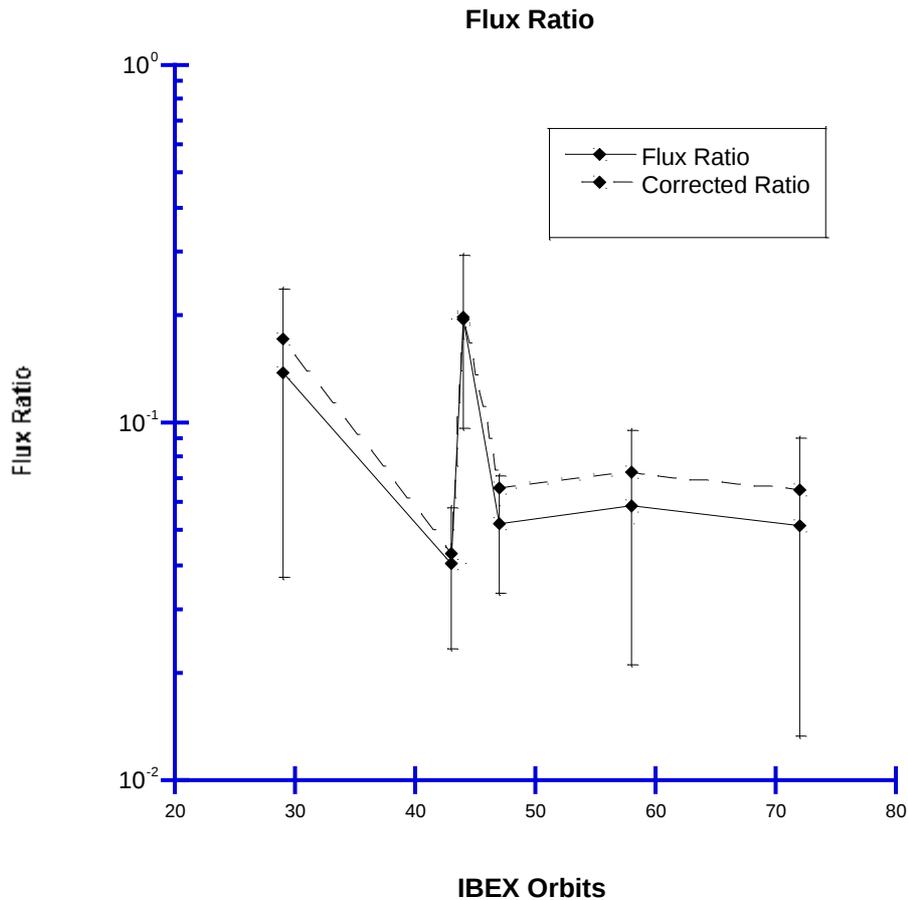

*Figure 2: Ratio of lunar ENA flux to impacting solar wind flux for the different orbits of IBEX-Lo. The error bars come from propagation of error as explained in [Rodriguez et al., 2012]. The dashed line represents the correction from the analysis in this work.*

igure 2: shows the original data for the fraction of reflected ENA flux, together with the corrected flux. We obtained the average of this corrected flux ratio (average global ENA albedo) over all orbits is 0.11±0.06.

**Conclusions**

We presented updated values of the reflectance of neutralized solar wind using the angular scattering function reported by Schaufelberger et al. (2011) and integrating over the viewing geometry of IBEX. The value for the reflectance of neutralized solar wind is $A_H = 0.11\pm0.06$, which is ~25% larger than the previous value of 0.09±0.05, based on an assumed uniform angular scattering distribution. A more thorough understanding of the scattering interaction would require an energy dependent scattering function. While the energy coverage of IBEX-LO and SARA does cover the relevant energy range of the reflected solar wind, the energy resolution of these instruments are not high enough to investigate such a scattering function. While an improved energy-dependant scattering function would help our understanding of the lunar regolith and its interaction with the solar wind, it would not substantially change the albedo determination. Further, as more ENA observations are made, a more thorough knowledge of the scattering function will be required to interpret and determine global properties of the objects in question.

The reflectance of neutralized solar wind is affected by a variety of factors, from bulk electromagnetic fields on the surface to material properties of space-weathered regolith. The present analysis has immediate relevance to the question of implantation rate of solar wind in the regolith, which is lower than assumed so far. The lowered hydrogen implantation could also affect the formation of the lunar hydrogen exosphere.

The proton implantation is an important contributor to the space weathering of the regolith, thus it is an important clue to understanding evolution of surfaces of solar system bodies. For example, it was suggested by Zeller et al. [1970] that water and other hydrocarbons might be formed from the solar wind protons via protolysis reactions. Gibson and Moore [1972] were the first to experimentally verify the formation of water using terrestrial olivine as lunar analogue, most recently such experiments were reported by Managadze et al. [2011].

There is also a possible useful feature of the lunar ENA emission for which the knowledge of the scattering function is prerequisite: as a "standard candle" for in-flight cross-calibration of ENA sensors. If the lunar ENA flux is known to within enough precision, future neutral sensors could use this as an in flight calibration mechanism, much as existing optical telescopes use well measured astronomical sources as calibration points. This work enables a more complete in-flight comparison of four existing ENA sensor datasets, the two on IBEX as well as two on the Chandrayaan spacecraft.